\documentclass[aps,prd,twocolumn,groupedaddress,nofootinbib]{revtex4}
\usepackage{bm}
\usepackage{graphicx}
\usepackage{times}
\usepackage{amssymb,amsmath}
\usepackage{rotating}
\usepackage{float}
\usepackage[position=b]{subfig}
\def\araa{ARAA}

\def\mnras{MNRAS}
\def\jcap{JCAP}

\def\apj{ApJ}
\def\apjl{ApJL}

\def\prd{PRD}

\def\beq{\begin{equation}}
\def\eeq{\end{equation}}
\def\bey{\begin{eqnarray}}
\def\eey{\end{eqnarray}}
\def\bfig{\begin{figure}}
\def\efig{\end{figure}}

\def\sun{\odot}
\def\lsim{\mathrel{\raise.3ex\hbox{$<$\kern-.75em\lower1ex\hbox{$\sim$}}}}
\def\gsim{\mathrel{\raise.3ex\hbox{$>$\kern-.75em\lower1ex\hbox{$\sim$}}}}

\begin{document}
\title{Cluster--Void Degeneracy Breaking: Modified Gravity in the Balance}

\author{Martin Sahl\'en$^{1,2}$}
\email{msahlen@msahlen.net}
\author{Joseph Silk$^{2,3,4,5}$}
\affiliation{Department of Physics and Astronomy, Uppsala University, SE-751 20 Uppsala, Sweden$^{1}$
\\
The Johns Hopkins University, Department of Physics \& Astronomy, 3400 N. Charles St., Baltimore, MD 21218, USA$^2$\\
Institut d'Astrophysique de Paris, 98 bis bd Arago, F-75014 Paris, France$^3$\\
AIM-Paris-Saclay, CEA/DSM/IRFU, CNRS, Universit\'e Paris 7, F-91191, Gif-sur-Yvette, France$^4$\\
BIPAC, University of Oxford, 1 Keble Road, Oxford OX1 3RH, UK$^5$
}

\begin{abstract}
Combining galaxy cluster and void abundances is a novel, powerful way to constrain deviations from General Relativity and the $\Lambda$CDM model. For a flat $w$CDM model with  growth of large-scale structure parameterized by the redshift-dependent growth index $\gamma(z) = \gamma_0 + \gamma_1 z/(1+z)$ of linear matter perturbations, combining void and cluster abundances in future surveys with {\it Euclid} and the 4-metre Multi-Object Spectroscopic Telescope (4MOST) could improve the Figure of Merit for ($w, \gamma_0, \gamma_1$) by a factor of 20 compared to individual abundances. In an ideal case, improvement on current cosmological data is a Figure of Merit factor $600$ or more. 
\end{abstract}


\maketitle

\begin{figure}[t]
\includegraphics[trim={0.15cm 0.1cm 1cm 0.15cm},clip,width=0.49\textwidth]{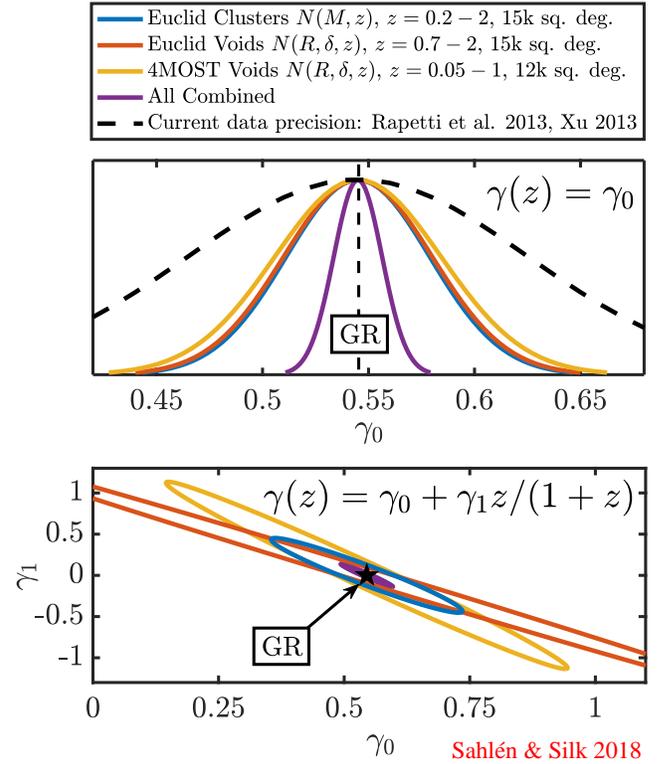}
\caption{ \label{fig:marg} Forecast marginalized PDFs for the growth index $\gamma(z)$ from cluster and void abundances in {\it Euclid} and 4MOST surveys, for a flat $w$CDM model with separate growth index $\gamma(z)$. Two different parameterizations are shown, with the two-parameter case $(\gamma_0, \gamma_1)$ displaying $68\%$ confidence contours. A PDF representative of current data precision \cite{2013MNRAS.432..973R, 2013PhRvD..88h4032X} is included as a dashed line. The fiducial value of $\gamma_0 = 0.545, \gamma_1 = 0$ for GR+$\Lambda$CDM is also indicated.}
\end{figure}

{\bf  INTRODUCTION}
Clusters and voids in the galaxy distribution are rare extremes of the cosmic web. As biased samples of the matter distribution, they can be used to place constraints on cosmological models. The abundances of clusters and voids are sensitive probes of dark energy \cite{2015arXiv150307690P, 2011ARA&A..49..409A,1992ApJ...388..234B}, modified gravity \cite{2015MNRAS.450.3319L,2011ARA&A..49..409A}, neutrino properties \cite{2010JCAP...09..014B,2015arXiv150603088M}, and non-Gaussianity \cite{2010ApJ...724..285C}.

In earlier work \cite{2016ApJ...820L...7S}, we derived the first statistically significant cosmological constraints from voids, showing that the joint existence of the largest known cluster and void strongly requires dark energy in the flat $\Lambda$CDM model. We also reported a powerful parameter complementarity between clusters and voids in the $\Lambda$CDM model. Here, we extend the modelling to the case where the dark energy equation of state and matter perturbation growth index are independent, free parameters. We investigate the complementarity between cluster and void abundances for constraining deviations from the General Relativity (GR)+$\Lambda$CDM model, and forecast ideal-case, prior-free constraints from future surveys.

{\bf FIDUCIAL SURVEYS}
We consider the {\it Euclid} Wide Survey \cite{2011arXiv1110.3193L} and the 4-metre Multi-Object Spectroscopic Telescope (4MOST) Galaxy Redshift Survey \cite{doi:10.1117/12.2055826}. Survey specifications are listed in Table~\ref{tab:surveys}. For voids, we limit ourselves to the spectroscopic segment of {\it Euclid} and the 4MOST spectroscopic survey, for which observational systematics should be relatively minimal (photometric redshifts can significantly distort the void shapes). We note that there is also a 4MOST cluster survey planned, which we do not consider here; our aim is to highlight the complementarity of clusters and voids, and of {\it Euclid} and 4MOST for void surveys. 

\begin{table}[htp]
\vspace{-5pt}
\caption{Fiducial survey specifications.}
\label{tab:surveys}
\begin{center}
\begin{tabular}{|c|c|c|c|}
\hline
Survey & Area [sq. deg.] & Redshift  \\
\hline
{\it Euclid} Clusters & $15000$ & $0.2 - 2.0$ \\
{\it Euclid} Voids & $15000$ & $0.7 - 2.0$  \\
4MOST Voids & $12000$ & $0.05-1$  \\ 
\hline
\end{tabular}
\end{center}
\vspace{-12pt}
\end{table}%

{\bf Cluster Selection and Limiting Cluster Mass}
The limiting cluster mass is chosen as $M_{200,c} = 8 \times 10^{13}\,h^{-1} M_{\sun}$ (where $M_{200,c}$ is the halo mass as defined by an overdensity of 200 above the critical density), with a constant $80\%$ completeness \cite{2015arXiv150502165S}. A constant completeness level is not exact, but sufficiently accurate for our forecasting purposes.

{\bf Void Selection and Limiting Void Radius}
We assume that void selection is complete for voids above the limiting radius $R_{\rm lim}$  (with radii defined in the galaxy field). The limiting radius is set by demanding that the void radius $R > 2 R_{\rm mps} = 2 \bar{n}_{\rm gal}^{-1/3}(z)$ \cite{2015arXiv150307690P}, where $\bar{n}_{\rm gal}(z)$ is the mean comoving galaxy number density. 

We use the following prescription for {\it Euclid}, which provides a good fit to the galaxy densities in \cite{2016arXiv160600180A}: 
\begin{eqnarray}
\nonumber
\frac{R_{\rm lim}(z)}{h^{-1}{\rm Mpc}} & = & 118.272 - 334.64z + 399.22z^2 - 207.26z^3 \\
 & & + 40.838z^4 \,.
\end{eqnarray}
For 4MOST, we assume that
\begin{eqnarray}
\frac{R_{\rm lim}(z)}{h^{-1}{\rm Mpc}} = \begin{cases}
13, & 0.05 \leq z \leq 0.5\\
31, &  0.5 < z \leq 0.7 \\
15,& 0.7 < z \leq 0.8 \\
17, & 0.8 < z \leq 0.9\\
42, & 0.9 < z \leq 1.0
\end{cases} \,,
\end{eqnarray}
based on the current survey plans \cite{4mostWP}.

{\bf Binning}
We use bins in redshift $\Delta z = 0.1$, cluster mass $\Delta \log (M_{200}) = 0.2$, void radius $\Delta \log (R) = 0.1$, and void density contrast $\Delta \delta_{\rm dm} = 0.3$ (from $-1$ up). This binning should accommodate expected measurement uncertainties.

{\bf MODEL} We predict cluster and void abundances adopting models and methodology developed in earlier work \cite{2016ApJ...820L...7S, Sahlen2009}. 

{\bf Cosmological Model}
We assume a flat $w$CDM background evolution. The primordial density perturbations follow a power-law power spectrum, and neutrinos are massless. The linear growth of perturbations is determined by the growth index $\gamma(a)$, with the linear growth rate given by \cite{2005PhRvD..72d3529L}
\begin{eqnarray}
f \equiv \frac{d \ln \delta}{ d \ln a} = \Omega^{\gamma(a)}_{\rm m}(a) \,,
\end{eqnarray}
where the scale factor $a = 1/(1+z)$. The model is specified by today's values of the Hubble parameter $h$, mean matter density $\Omega_{\rm m}$, dark energy equation of state $w$, mean baryonic matter density $\Omega_{\rm b}$, statistical spread of the matter field at quasi-linear scales $\sigma_8$, scalar spectral index $n_{\rm s}$, and growth index $\gamma(z)$. We consider i) $\gamma(z) = \gamma_0 + \gamma_1z/(1+z)$, ii) $\gamma(z) = \gamma_0$ \cite[e.g.][]{DiPorto:2012ey}.

{\bf Number Count Model}
We model cluster and void number counts as in \cite{2016ApJ...820L...7S}, but with the growth of linear perturbations described by growth index $\gamma(a)$, and background by a flat $w$CDM model with a constant dark-energy equation of state $w$. 

The abundance model is given by 
\begin{equation}
\label{eq:numbercount}
\bar{N} = \int \int \int p(O | O_{\rm t}) n[M(O_{\rm t}), z]  \frac{dM}{dO_{\rm t}} \frac{dV}{dz} dz dO_{\rm t} dO,
\end{equation}
where $O$ is the observable (mass, radius) for clusters or voids, $O_{\rm t}$ the true physical value of the observable $O$, and $M(O_{\rm t})$ the (unbiased) mass estimate of the object. The differential number density is $n(M,z)$, $p(O | O_{\rm t})$ is the probability density function (PDF) of assigning an observed value $O$ for a true value $O_{\rm t}$, and $dV/dz$ is the cosmic volume element. For integrating Eq.~(\ref{eq:numbercount}), we use $M_{\rm void}=\frac{4}{3}\pi R^3\rho_{\rm m}(1+\delta_{\rm dm})$.

{\bf Number Density}
The differential number density of objects in a mass interval ${\rm d}M$ about $M$ at redshift $z$ is
\begin{equation}
\label{eq:numdens}
n(M, z)\,{\rm d}M = -F(\sigma, z)\,\frac{\rho_{\rm m}(z)}{M\sigma(M, z)}\,
\frac{{\rm d}\sigma(M, z)}{{\rm d}M}\,{\rm d}M\,,
\end{equation}
where $\sigma(M, z)$ is the dispersion of the density field at some comoving scale $R_L=(3M/4\pi\rho_{\rm m})^{1/3}$, and
$\rho_{\rm m}(z) = \rho_{\rm m}(z=0)(1+z)^3$ the matter density. The expression can be written in terms of linear-theory radius $R_L$ for voids. The multiplicity function (MF) denoted $F(\sigma, z)$ is described in the following for clusters and voids. 

{\bf Cluster MF} 
The cluster (halo) MF $F_h(\sigma)$ encodes halo collapse statistics. We use the MF of \citet{Watson2013}, their Eqs.~(12)-(15). Mass conversions are performed using the methods in \cite[Appendix C]{2003ApJ...584..702H}.

{\bf Void MF} 
We employ the simulation-calibrated void MF in \cite{2015arXiv150307690P} based on a Sheth--van de Weygaert form \cite{Sheth2004}, 
\begin{equation}
\label{eq:fv}
F_{\rm v} = \sqrt[]{\frac{2\nu}{\pi}}e^{-\nu/2}\,,
\end{equation}
where $\nu = \delta^2_{\rm v} / \sigma^2(R_{\rm L},z)$, and for which a critical density threshold $\delta_{\rm v} = -0.45$ was derived for shell-crossed voids. We find that void-in-cloud corrections \cite{Sheth2004} are negligible for our analysis, so neglected those in Eq.~(\ref{eq:fv}). We generalize this prescription to other density contrasts through the spherical-expansion relationship \cite{Jennings2013}
\begin{equation}
\label{eq:sphexp}
\delta_{\rm v} = c[1-(1+b_{\rm eff}^{-1}\delta_{\rm dm})^{-1/c}] \,,
\end{equation}
where $c=1.594$, and we have set $\delta_{\rm m} = b_{\rm eff}^{-1}\delta_{\rm dm}$ (which also defines $b_{\rm eff}$). For the calibration in \cite{2015arXiv150307690P} we have $\delta_{\rm v} = -0.45$ and $\delta_{\rm dm} = -0.8$, which yields $b_{\rm eff} \approx 2.44$. We then use Eq.~(\ref{eq:sphexp}) for other values of the dark-matter density contrast $\delta_{\rm dm}$, to convert to a linear density contrast $\delta_{\rm v}$ to be used as the corresponding density threshold in the void MF.  The void MF for (non-linear) radius $R$ is evaluated at corresponding linear radius $R_{\rm L}$, which here is related as $R/R_{\rm L} = (1+\delta_{\rm dm})^{-1/3}$. Note that these spherical-expansion dynamics do not include any dark-energy or modified-gravity effects, but such corrections are sub-dominant for the model we consider \cite{Jennings2013}. 

While our prescription for generalizing the void MF to general density contrasts should in principle be calibrated with full simulations of the galaxy field, it is robust with respect to our conclusions (e.g., we have tested the effect of varying the value of the bias, and of a bias defined on the linear density field).

{\bf Scatter}
We include scatter in cluster and void properties as log-normal PDFs $p(O | O_{\rm t})$ for the observable $O$ (i.e. $M_{200}$ or $R$) given its true value $O_{\rm t}$.  
The intrinsic scatter between observed and true cluster mass is given by \cite{2015arXiv150502165S}
\beq
\sigma^2_{\ln M(z)} = \sigma^2_{\ln M, 0} - 1 + (1+z)^{2\beta}
\eeq
with $\sigma^2_{\ln M, 0} = 0.2$, $\beta = 0.125$, based on $N$-body simulation results. The intrinsic scatter between observed and true (spherical-equivalent) void radius is not well-studied. We assume that
\beq
\sigma^2_{\ln R(z)} = \sigma^2_{\ln R, 0}
\eeq
with $\sigma^2_{\ln R, 0} = 0.2$, which is a reasonable first approximation given that e.g. ellipticity varies but typically is of the order $15\%$ \citep{2015JCAP...03..047L}.

{\bf Fiducial Parameters}
We assume $h=0.7$,  $\Omega_{\rm m} = 0.3$, $\gamma_0 = 0.545$, $\gamma_1 = 0$, $\Omega_{\rm b} = 0.045$, $\sigma_8=0.8$, $w=-1$, $n_{\rm s}=0.96$, $\Sigma m_{\nu} = 0 \,{\rm eV}$, and three neutrino species so that the early-universe effective relativistic degrees of freedom $N_{\rm eff}~=~3.046$.

{\bf LIKELIHOOD}
We model the number counts of clusters and voids as Poisson-distributed in each bin, and bins to be statistically independent. Hence, the log-likelihood is
\beq
\ln \mathcal{L} = \sum_i N_i \ln \bar{N}_i - \bar{N}_i  \,, 
\eeq
where $N_i$ is the observed number of objects in bin $i$, and $\bar{N}_i$ is the model prediction, Eq.~(\ref{eq:numbercount}), for the expected number of objects in the same bin.

{\bf COMPUTATION}
We compute a Fisher matrix estimate of expected parameter constraints based on the Poisson likelihood \cite{2015arXiv150307690P}. This leads to a Fisher matrix 
\beq
\mathcal{F}_{mn} = \sum_i \frac{1}{\bar{N}_i} \frac{\partial \bar{N}_i}{\partial \theta_m} \frac{\partial \bar{N}_i}{\partial \theta_n} \,,
\eeq 
where $\bar{N}_i$ is the fiducial expected number of objects in bin $i$ and $\theta_m$ are the different model parameters under consideration. The corresponding covariance matrix $\mathcal{C} = \mathcal{F}^{-1}$. The space of 8 free parameters is defined by $\{\Omega_{\rm m}, \gamma_0, \gamma_1, w,  \sigma_8,  n_{\rm s}, h, \Omega_{\rm b} \}$, and we also consider the 7-parameter case where $\gamma_1 = 0$. 
Background evolution and linear power spectrum computations are performed using a modified version of CAMB \cite{1996ApJ...469..437S}. 

\begin{table}[htp]
\vspace{-5pt}
\caption{Forecast parameter uncertainties for growth-index model $\gamma(z) = \gamma_0 + \gamma_1z/(1+z)$.}
\label{tab:constrz}
\begin{center}
\begin{tabular}{|c|c|c|c|c|c|c|c|c|c|c|}
\hline
Survey & $\sigma(\Omega_{\rm m})$ & $\sigma(\gamma_0)$ & $\sigma(\gamma_1)$ & $\sigma(w)$ & $\sigma(\sigma_8)$ & $\sigma(n_{\rm s})$ & $\sigma(h)$ & $\sigma(\Omega_{\rm b})$ \\
\hline
EC\footnote{{\it Euclid} Clusters} & $0.002$ & $0.13$ & $0.30$ & $0.006$ & $0.01$ & $0.09$ & $0.10$ & $0.010$\\
EV\footnote{{\it Euclid} Voids} & $0.005$ & $0.53$ & $0.97$ & $0.01$ & $0.07$ & $0.03$ & $0.02$ & $0.003$ \\
4V\footnote{4MOST Voids} & $0.002$ & $0.26$  & $0.75$ & $0.01$ & $0.02$ & $0.04$ & $0.03$ & $0.003$ \\ 
EV+4V & $0.002$ & $0.12$  & $0.25$ & $0.005$ & $0.01$ & $0.02$ & $0.02$ & $0.002$ \\ 
All & $0.0006$ & $0.03$  & $0.09$ & $0.003$ & $0.003$ & $0.01$ & $0.01$ & $0.002$ \\ 
\hline
\end{tabular}
\end{center}
\vspace{-12pt}
\end{table}%

\begin{table}[htp]
\vspace{-5pt}
\caption{Forecast and current parameter uncertainties for growth-index model $\gamma(z) = \gamma_0$.}
\label{tab:constr}
\begin{center}
\begin{tabular}{|c|c|c|c|c|c|c|c|c|c|}
\hline
Survey  & $\sigma(\Omega_{\rm m})$ & $\sigma(\gamma_0)$ & $\sigma(w)$ & $\sigma(\sigma_8)$ & $\sigma(n_{\rm s})$ & $\sigma(h)$ & $\sigma(\Omega_{\rm b})$ \\
\hline
EC & $0.001$ & $0.03$ & $0.003$ & $0.008$ & $0.07$ & $0.08$ & $0.009$ \\
EV & $0.005$ & $0.04$ & $0.01$ & $0.01$ & $0.03$ & $0.02$ & $0.003$ \\
4V & $0.002$ & $0.04$  & $0.01$ & $0.01$ & $0.04$ & $0.03$ & $0.003$ \\ 
EV+4V & $0.002$ & $0.02$  & $0.005$ & $0.005$ & $0.02$ & $0.02$ & $0.002$ \\ 
All & $0.0006$ & $0.01$  & $0.002$ & $0.003$ & $0.01$ & $0.01$ & $0.001$ \\ 
\hline
Current \cite{2013MNRAS.432..973R, 2013PhRvD..88h4032X} & $0.01$ & $0.08$ & $0.05$ & $0.02$ & $0.006$ & $0.01$ & $0.001$ \\
\hline
\end{tabular}
\end{center}
\vspace{-12pt}
\end{table}%

\begin{table}[htp]
\vspace{-5pt}
\caption{Forecast and current Figures of Merit (FoM) for the dark-energy and modified-gravity parameters $w$, $\gamma_0$ and $\gamma_1$ in the two growth-index cases considered.}
\label{tab:fom}
\begin{center}
\begin{tabular}{|c|c|c|}
\hline
Survey & FoM ($w, \gamma_0$) & FoM ($w, \gamma_0, \gamma_1$) \\
\hline
EC &  $1.1 \times 10^4$ & $3.4  \times 10^4$ \\ 
EV & $2.6\times 10^3$ & $2.6  \times 10^3$ \\
4V & $3.0\times 10^3$ &  $3.8  \times 10^3$\\
EV+4V &  $1.7 \times 10^3$ & $4.2  \times 10^4$ \\
All & $3.9 \times 10^4$ &  $6.6  \times 10^5$\\
\hline
Current  & $\sim 300$ \cite{2013MNRAS.432..973R, 2013PhRvD..88h4032X} & $\mathcal{O}(10^2-10^3)$\footnote{Estimated based on using $\sigma(w), \sigma(\gamma_0)$ in 
\cite{2013MNRAS.432..973R} and $\sigma(\gamma_1)$ in \cite{2017PhRvD..96f3517B}. These are upper limits on current uncertainties. These are then scaled up according to the factor differences between Tables \ref{tab:constrz}~\&~\ref{tab:constr}. The estimate is robust with respect to parameter correlations.} \\ 
\hline
\end{tabular}
\end{center}
\vspace{-12pt}
\end{table}%

{\bf RESULTS} 
{\bf Expected Numbers}
We predict $5 \times 10^5$ clusters and $9 \times 10^5$ voids in the {\it Euclid} cluster and void surveys, and $4 \times 10^5$ voids in the 4MOST void survey. These numbers are consistent with earlier predictions \cite{2015arXiv150502165S, 2015arXiv150307690P}.

\onecolumngrid 
\begin{center}
\begin{sidewaysfigure*}[h]
\includegraphics[trim={3.8cm 1.1cm 3.6cm 1.8cm},clip,width=\textwidth]{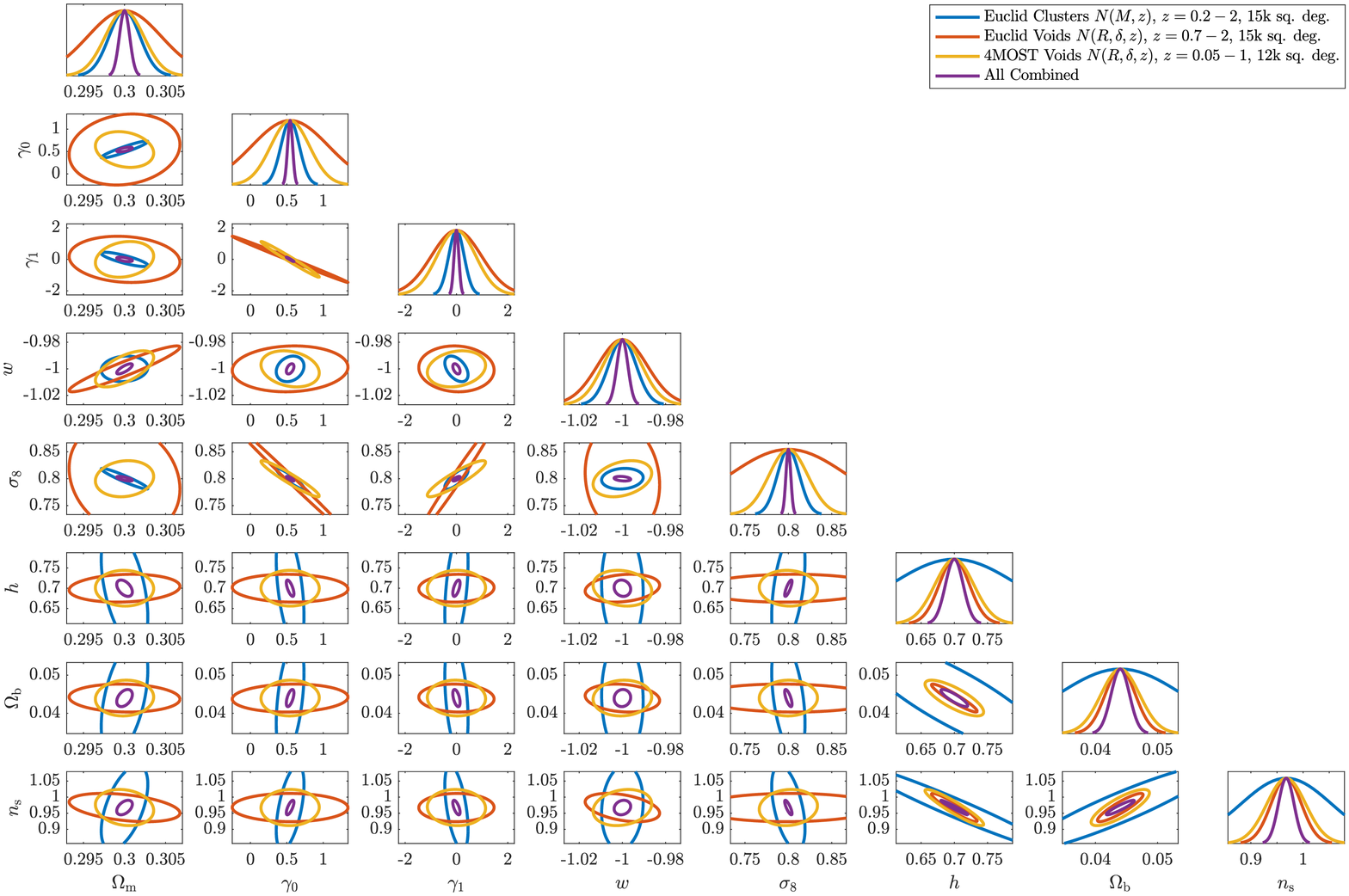}
\caption{ \label{omms8} Forecast $68\%$ parameter contours, and marginal PDFs, from cluster and void abundances in future {\it Euclid} and 4MOST surveys, for a flat $w$CDM model with separate growth index $\gamma(z) = \gamma_0 + \gamma_1 z/(1+z)$. }
\end{sidewaysfigure*}
\end{center}
\twocolumngrid

{\bf Parameter Constraints}
The forecast ideal-case parameter constraints are shown in Fig.~\ref{fig:marg} (marginalized constraints on $\gamma_0, \gamma_1$ for both growth-index models) and Fig.~\ref{omms8} (complete set of 1-D and 2-D PDFs, redshift-dependent growth index only) for the separate data sets and combinations thereof. Table~\ref{tab:constrz}~\&~\ref{tab:constr} list the forecast marginalized parameter uncertainties. In this ideal scenario, all surveys can improve substantially on current parameter uncertainties except for $h$, $\Omega_{\rm b}$ and $n_{\rm s}$. Hence, including data more informative on these parameters (e.g. cosmic microwave background data) will be a valuable, but here not crucial, addition. 

{\bf Figures of Merit}
The Figures of Merit (FoM) for the dark-energy and modified-gravity parameters, defined by
\begin{equation}
{\rm FoM} = \frac{1}{\sqrt{\det {\rm cov}({\rm parameters})}} \,,
\end{equation}
are listed in Table~\ref{tab:fom}. Compared to current data, an improvement factor $\mathcal{O}(10^2-10^3)$ is expected for both growth-index models in this ideal case. The {\it Euclid} cluster survey is at least as informative (roughly) as are the combined {\it Euclid} + 4MOST void surveys. When all surveys are combined, a factor $\sim 4-20$ improvement in FoM is seen in comparison. With more detailed modelling of the galaxy cluster and void distributions including mass-observable scaling relations and other sources of uncertainty, the relative improvement can be expected to be greater, since such physics and systematics are mostly independent between clusters and voids.  The details of this, particularly for voids, is the subject for ongoing work in the field.

{\bf Parameter Sensitivity} 
{\it Clusters} Cluster sensitivity to cosmological parameters is described extensively in the literature \cite[e.g.][]{2002ApJ...577..569L,2011ARA&A..49..409A}.

{\it Voids: General}
To examine the sensitivity of void abundance to cosmological parameters, we consider  the redshift-dependent growth index model and a generic survey with a fixed limiting void radius $R_{\rm lim} = 10\,h^{-1}{\rm Mpc}$ and galaxy bias akin to {\it Euclid} and 4MOST. The survey is assumed large enough to give an unbiased sample of the largest voids (sky coverage $f_{\rm sky} \gtrsim 0.1$), and $z=0.05 - 2.05$.

\begin{center}
\begin{sidewaysfigure*}[h]
	\subfloat[Deep voids ($\delta_{\rm dm}= -0.85$).]{\includegraphics[trim={1.1cm 1.1cm 1.0cm 0.4cm},clip,width=0.5\textwidth]{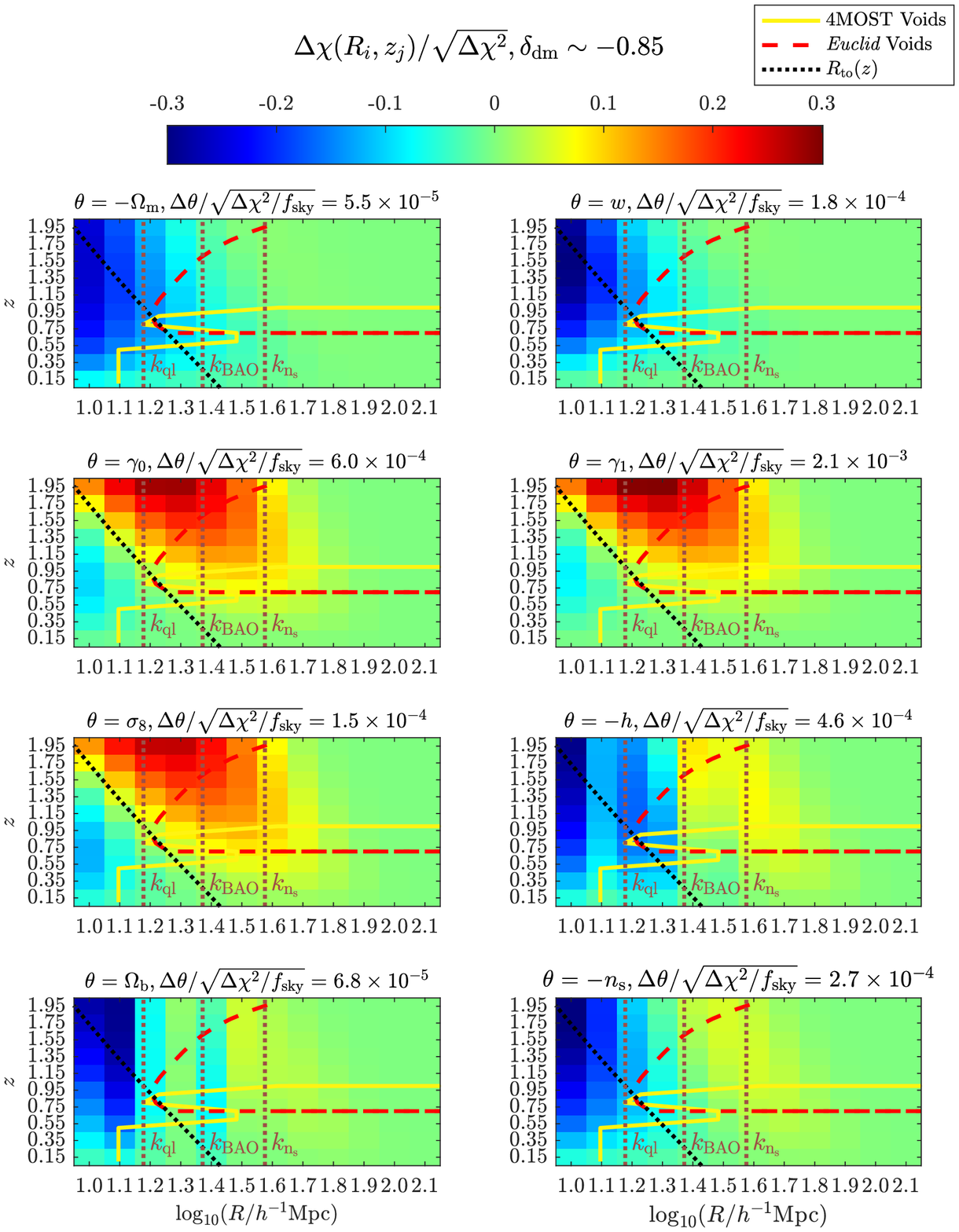}}
	\subfloat[Shallow voids ($\delta_{\rm dm}= -0.25$).]{\includegraphics[trim={1.1cm 1.1cm 1.0cm 0.4cm},clip,width=0.5\textwidth]{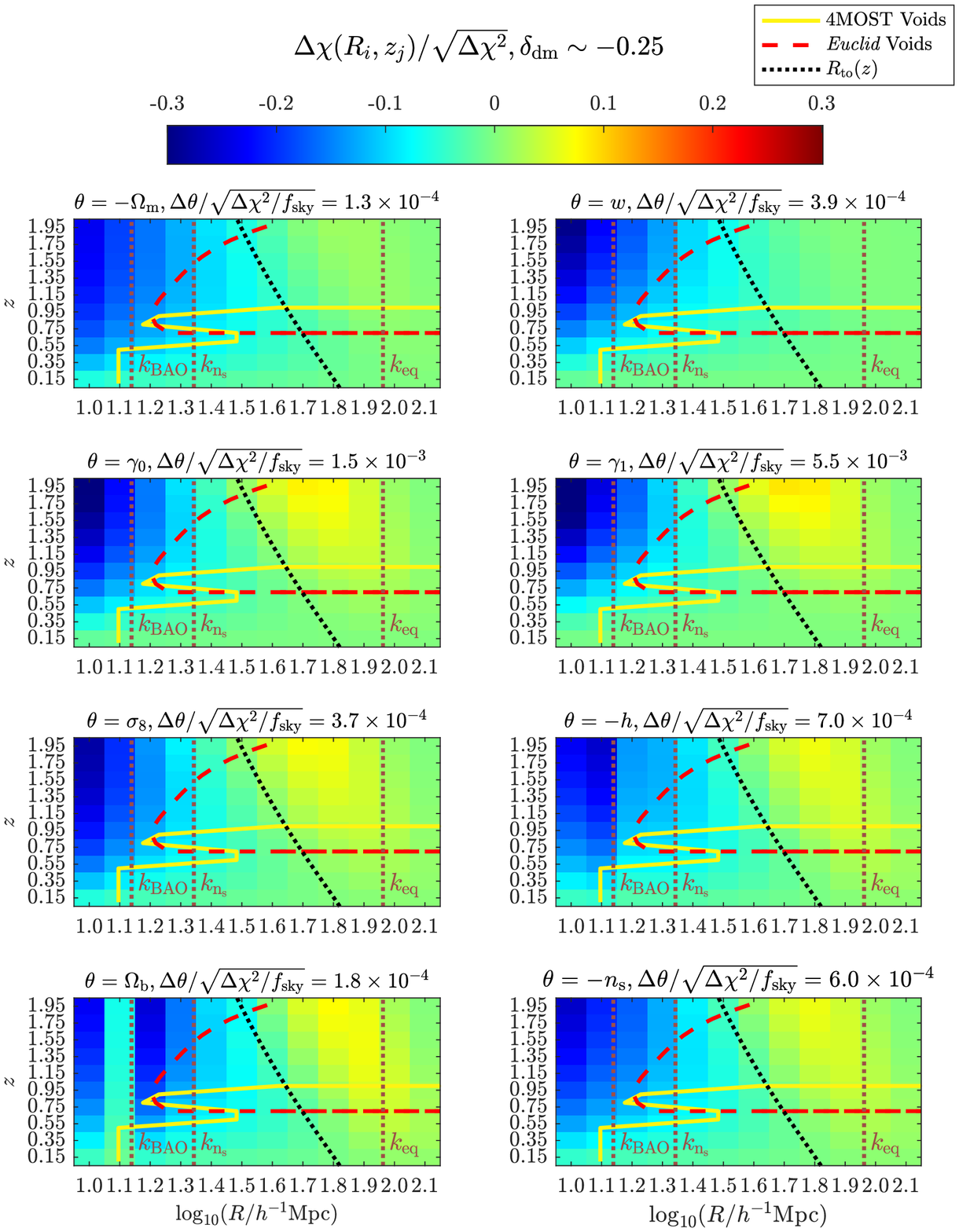}}
\caption{ \label{fig:distshift} Void distribution parameter sensitivity in the redshift-dependent growth index model. We assume the same fiducial cosmological and galaxy bias models as elsewhere, but consider a generic survey of deep voids with $R_{\rm lim} = 10\,h^{-1}{\rm Mpc}$ and $z=0.05 - 2.05, \Delta \log_{10} (R/h^{-1}Mpc) = 0.1, \Delta z = 0.2$. For each parameter, the figure shows $\Delta \chi^{\rm rel}_{i,j}$ when that parameter {\it only} is varied (hence, $\sigma_8$ is kept normalized to the fiducial value at $z=0$ when other parameters are varied). The indicated turn-over radius $R_{\rm to}(z)$ (black, dotted lines) is clearly visible in the sensitivity to power-spectrum and growth parameters. Scales related to the cosmological parameters are in brown, dotted lines ($k_{\rm ql} = 0.13h$ Mpc$^{-1}$, $k_{\rm BAO} = 0.06h$ Mpc$^{-1}$, $k_{\rm n_s} = 0.05h$ Mpc$^{-1}$, $k_{\rm eq} = 0.012h$ Mpc$^{-1}$). Note the distinct baryon acoustic oscillation signature in the $\Omega_{\rm b}$ panels. The radius-redshift coverage of the surveys are also shown (4MOST, yellow solid lines; {\it Euclid}, red dashed lines). Some $\theta$'s are defined with negative sign to aid comparison. Note also that $\Delta \chi^{\rm rel}_{i,j}$ does \emph{not} depend on $f_{\rm sky}$. The normalization $\Delta \theta/\sqrt{\Delta \chi^2/f_{\rm sky}}$ is also given for each parameter. Quantities are described further in the text.}
\end{sidewaysfigure*}
\end{center}

The effects of variations in cosmological parameters on void abundance, and their statistical significance, are shown in Fig.~\ref{fig:distshift}. The limiting radii of the {\it Euclid} and 4MOST surveys are also indicated in the figure. We discuss deep voids ($\delta_{\rm dm} \sim -0.85$) only, but the translation to medium-deep ($\delta_{\rm dm} \sim -0.55$) or shallow ($\delta_{\rm dm} \sim -0.25$) voids is straightforward, as Fig.~\ref{fig:distshift} indicates. We illustrate the void sensitivity to cosmological parameters with the relative change
\begin{eqnarray}
\Delta \chi^{{\rm rel}}_{i,j}(\Delta \theta_k) \equiv \frac{\Delta \chi(R_i, z_j; \Delta \theta_k, f_{\rm sky})}{\sqrt{\Delta \chi^2(\Delta \theta_k, f_{\rm sky})}}
\end{eqnarray}
for some small, positive single-parameter change $\Delta \theta_k$ away from the fiducial model ($k$ indexes the cosmological parameters). Here, 
\begin{eqnarray}
\nonumber
\Delta \chi(R_i, z_j; \Delta \theta_k, f_{\rm sky}) \equiv \\
\frac{\bar{N}_{i,j}(\theta_k + \Delta \theta_k, f_{\rm sky})-\bar{N}_{i,j}(\theta_k, f_{\rm sky})}{\sqrt{\bar{N}_{i,j}(\theta_k, f_{\rm sky})}}
\end{eqnarray}
is the number count change in bin $i, j$ relative to the fiducial model, in units of the corresponding Poisson uncertainty; and $\Delta \chi^2(\Delta \theta_k, f_{\rm sky}) = \Sigma_{i,j} \Delta \chi^2(R_i, z_j; \Delta \theta_k, f_{\rm sky})$ is the total (Poisson) $\Delta \chi^2$ across all bins. 

For small $\Delta \theta_k$, we approximate 
\begin{eqnarray}
\Delta \chi(R_i, z_j; \Delta \theta_k, f_{\rm sky}) = 
\sqrt{\frac{2f_{\rm sky}}{\bar{N}_{i, j}}}\frac{\partial \bar{N}_{i,j}}{\partial \theta_k} \Delta \theta_k \,.
\end{eqnarray}
Then, $\Delta \chi^{{\rm rel}}_{i,j}$ is independent of  $f_{\rm sky}$ and $\Delta \theta_k$ (normalizing to the same total $\Delta \chi^2$ implicitly picks some set of $\Delta \theta$'s which all separately produce the same total $\Delta \chi^2$). 

The results in Fig.~\ref{fig:distshift} show a few general features: i) suppression of small voids, ii) relative enhancement of large voids, iii) a redshift-dependent turnover between suppression and enhancement,  
iv) variation in scale-dependence. (Suppression and enhancement switch with change of sign in $\Delta \theta$.) We discuss these features in the following. 

i) Small-scale suppression is generically produced by changes of the comoving volume. Variations in $\Omega_{\rm m}$ and $w$ will have this effect. 

ii) Large-scale enhancement is usually accompanied by small-scale suppression. This is because these variations are all due to changes in the matter-field dispersion $\sigma(R,z)$. A positive shift $\Delta \sigma$ (due to variation in parameters affecting power spectrum or growth) effectively changes the curvature of the void MF, such that small scales are suppressed and large scales enhanced. This follows from noting that
\begin{equation}
\label{eq:dlnfv}
\frac{d\ln F_{\rm v}}{d \ln \sigma} = \nu - 1\,,
\end{equation}
where $\nu \equiv \delta^2_{\rm v}/\sigma^2(R,z)$, so small/common voids ($\nu < 1$) are suppressed and large/rare voids ($\nu > 1$) enhanced. 

iii) The redshift-dependent turnover scale between suppression and enhancement is also explained by Eq.~(\ref{eq:dlnfv}). For voids with $\nu(R,z) = 1$, $\Delta F_{\rm v} / F_{\rm v} \approx 0$. The turnover scale $R_{\rm to}$, defined by the equation
\begin{equation}
\sigma(R_{\rm to},z) = |\delta_{\rm v}|\,,
\end{equation}
is fairly insensitive to variations in cosmological parameters. For shallow voids $R_{\rm to}(z=0) \sim 65 \,h^{-1}$Mpc, medium-deep voids $R_{\rm to}(z=0) \sim 35 \,h^{-1}$Mpc, and deep voids $R_{\rm to}(z=0) \sim 25\,h^{-1}$Mpc. (Note that $R_{\rm to}$ depends on survey characteristics, e.g. galaxy bias.)

iv) variation in scale dependence of suppression/enhancement arises primarily due to different parameters having different effects on the small-scale matter power spectrum. Small voids ($R_{\rm L} \lesssim 25\,h^{-1}$Mpc) are sensitive to the baryon acoustic peaks, and hence both shifts in scale ($\Omega_{\rm m}h$) and power suppression ($\Omega_{\rm b}h^2, n_{\rm s}$) will distinctly impact the void distribution. Large voids ($R_{\rm L} \gtrsim 90\,h^{-1}$Mpc) are also sensitive to the turnover scale of the matter power spectrum set by matter--radiation equality ($\Omega_{\rm m}h$). In addition, the relative importance of comoving volume vs. density-field statistics may also play a role. 

Thanks to tracer bias and the non-linear evolution of voids, a particular tracer-defined void radius $R_{\rm tr}$ will correspond to a linear comoving scale 
\begin{equation}
R_{\rm L} \approx R_{\rm tr}\frac{(1+b_{\rm tr}^{-1}\delta_{\rm tr})^{1/3}}{\beta(b_{\rm tr})} \gtrsim \frac{R_{\rm tr}}{2\beta(b_{\rm tr})}\,,
\end{equation}
where $b_{\rm tr}$ and $\delta_{\rm tr}$ are the tracer-defined bias and density contrast, and $\beta(b_{\rm tr}) \equiv R_{\rm tr}/R$ relates the tracer and matter-field radii. Thus, the deepest voids correspond to the smallest scales. For $b_{\rm tr}>1$, $\beta(b_{\rm tr})>1$, e.g. $\beta(1.4) \approx 1.2$ \cite{2016ApJ...820L...7S}. \emph{Hence, a survey of deep voids can probe linear comoving scales up to a factor $\sim 2-4$ smaller than the limiting void radius of the survey.}

For medium-deep and shallow voids, the $\Delta \chi^{{\rm rel}}_{i,j}$ patterns are qualitatively the same as in Fig.~\ref{fig:distshift}, but with scales shifted a factor $1.9$ and $2.6$ respectively. 

Putting these considerations together, we suggest the following conceptual picture of how void counts constrain parameters. The {\it relative} numbers of deep and shallow voids at a given redshift give an effective measure of the matter power spectrum on small scales relative to large scales (i.e. its shape) -- independent of growth or volume. Since the characteristic scale of the void samples change with redshift, and deep and shallow voids probe different scales in the primordial power spectrum, a wide range of scales can be constrained. Some relevant scales are indicated in Fig.~\ref{fig:distshift}, also showing the sensitivity to baryon acoustic oscillations. The characteristic void scale is a direct measure of the turnover radius $R_{\rm to}(z)$, so its evolution additionally measures $\sigma(R_{\rm to},z)$. The {\it absolute} number counts can then measure the background expansion via the direct effect of $\Omega_{\rm m}$ and $w$ on cosmic volume and their indirect effect on the growth rate. The turnover radius roughly defines the boundary between volume-dominated and growth-dominated voids. The growth-dominated counts also additionally constrain $\sigma(R,z)$. This picture suggests that void counts can constrain the background expansion, shape of the power spectrum, and growth history independently.

\textbf{\it Voids: Parameters with {Euclid} + 4MOST}
We find that deep voids provide the strongest parameter constraints, except for $\Omega_{\rm m}$ and $w$ with {\it Euclid}, where shallow voids do best. Looking at Fig.~\ref{fig:distshift} this is not surprising, since the sensitivity within the {\it Euclid} region is greater for shallow voids. However, shallow-void parameter constraints are also similar and complementary to deep voids, such that the combined constraints are tighter than the individual ones. An exception to this is $\gamma_0, \gamma_1$ with 4MOST, where the deep voids provide almost all constraining power. 

Looking at Tables~\ref{tab:constrz}~\&~\ref{tab:constr}, there is a difference in constraining power between the one-parameter and two-parameter growth models only in the normalization and redshift evolution of the power spectrum ($\sigma_8, \gamma_0, \gamma_1$). This agrees well with the expectation that background expansion, power spectrum shape and growth history can be independently constrained. Indeed, the dominant degeneracy is contained within $\sigma_8(z)$.

The 4MOST survey constrains $\Omega_{\rm m}$ better than {\it Euclid}. This derives from 4MOST containing deep voids smaller than the turnover radius (see Fig.~\ref{fig:distshift}). Such voids are sensitive to the growth of cosmic volume, not just growth of structure (as larger voids predominantly are). This produces degeneracy directions between $\Omega_{\rm m}$ and $w$ which rotate with redshift up to $z \sim 0.8$ (where the turnover radius exits the survey, and volume growth slows down) and settle on the growth-dominated degeneracy seen in Fig.~\ref{omms8}. The successively rotated degeneracies, when combined, constrain $\Omega_{\rm m}$ better than the growth-only constraints obtained with {\it Euclid}.

Differences in redshift sensitivity explains why {\it Euclid} and 4MOST void constraints are complementary (Fig.~\ref{omms8}), due to different redshift coverage (despite the {\it Euclid} void sample being twice as large). Specifically, the redshift evolution of $\sigma(R,z)$ across the survey is much weaker in {\it Euclid} ($5\%$) than in 4MOST ($20\%$) and $\Omega_{\rm m}(z)$ gets close to $1$. This implies that sensitivity to this redshift evolution will be correspondingly weaker. Uncertainties on $\sigma_8, \gamma_0, \gamma_1$ should then scale roughly as $\sqrt{N_{\rm 4V}/N_{\rm EV}}[\Delta \sigma_{\rm 4V}/\Delta \sigma_{\rm EV}] \approx 2.7$ between {\it Euclid} and 4MOST (neglecting parameter correlations). This agrees well with Table~\ref{tab:constrz}, where {\it Euclid} void uncertainties are $\sim 1.3-3.5$ times the 4MOST uncertainties on those parameters. In the constant growth-index model, this difference largely disappears thanks to breaking the $\gamma_0$--$\gamma_1$ degeneracy by setting $\gamma_1 = 0$. 

Medium-deep void counts add marginal additional information relative to shallow + deep void counts (adding them produces at most a $25\%$ reduction in the standard deviation of any parameter), but may be useful for calibration/systematics or tests of scale-dependent features. The results are consistent with the finding that most voids in the Baryon Oscillation Spectroscopic Survey (BOSS) have a density contrast minimum between $-0.9$ and $-0.6$, and radius between $20$ and $40$~$h^{-1}$ Mpc \cite{2016MNRAS.461..358N}. Voids that fall outside these ranges are relatively rare, with great statistical weight. Consequently, the intermediate density-contrast bin adds relatively little constraining power compared to the deep and shallow bins. The BOSS analysis, finding a $3\sigma$ discrepancy in the number of deep voids relative to the simplest allowed $\Lambda$CDM model, also independently hints that the void density-contrast distribution contains novel cosmological information. 

{\it Cluster--Void Complementarity}
Cluster and void number count parameter constraints are complementary for several parameters. In the redshift-dependent growth-index model, $\gamma_0$ and $\gamma_1$ are strongly correlated (Pearson correlation coefficient $\rho \sim -0.98$) regardless of data combination. However, $w$ has varying correlation with $\gamma_0, \gamma_1$ in the different surveys. In the {\it Euclid} void survey, the correlations are $\rho \sim -0.07$ to $0.07$. For all other survey combinations, the correlations vary between $|\rho| \sim 0.2 - 0.5$, but complementarity still reduces the overall uncertainty. 

In the constant growth-index model, the parameters $w$ and $\gamma_0$ are weakly correlated. For the individual surveys, $\rho \sim -0.28$ to $0.28$. For the combined {\it Euclid} + 4MOST void surveys, $\rho = -0.09$, and for all surveys combined $\rho = 0.13$. For current data constraints, $\rho \sim -0.6$ \cite{2013MNRAS.432..973R, 2013PhRvD..88h4032X}.

Complementary parameter degeneracies arise between clusters and voids, because they have different sensitivity to structure growth vs. volume expansion with redshift, comoving linear scales, and orthogonal sensitivities between matter disperson $\sigma$ and $\Omega_{\rm m}$ \cite[Section 4.5]{2016ApJ...820L...7S}. While {\it Euclid} clusters are better, separately, at constraining structure growth, the void samples are better at constraining the shape of the matter power spectrum.

 {\bf Systematics}
We do not explicitly marginalize over any systematics, but have included a net effect on number counts through statistical scatter in cluster masses and void radii. The value of this scatter is assumed to be known in the forecasts, since our purpose is to establish an ideal-case limit. In the case of voids, the expected value of this scatter is not well-known, but we consider only spectroscopic data to limit photometric shape distortions. It could arise due to e.g. intrinsic ellipticity, projection and Alcock--Paczynski effects, and redshift-space distortion. We also expect linear voids to have more irregular shapes than non-linear voids, so scatter should vary with density contrast and redshift. The impact of these effects is a subject for further study; some also contain additional cosmological information. Since shallow + deep voids contain most of the cosmological information of a void-count survey, medium-deep voids could potentially be used to self-calibrate void surveys.

Our void MF is a rough approximation, suited to this proof of concept. Accurate theoretical predictions based on large-scale simulations including non-linear modified gravity effects, detailed void characteristics, selection methods, and survey specifications are required for detailed forecasts and future real analyses. The detailed completeness in $R$ and $\delta$, and sources of bias such as survey boundary effects \cite{2014MNRAS.442.3127S, 2016MNRAS.461..358N}, all require further study.

Cluster samples can suffer bias due to poor mass calibration and scaling relations, skewed redshift estimates, poorly understood selection, or MF modelling, but these issues are not expected to prevent percent-level cluster cosmology with e.g. {\it Euclid} \cite{2015arXiv150502165S}.

Ultimately, combining clusters and voids (in conjunction also with e.g. cosmic microwave background data) will help limit the impact of systematics since they, as shown here, are relatively independent probes.

{\bf CONCLUSION}
We find that shallow + deep voids contain almost all the cosmological information of void counts, unless models with e.g. additional scale dependence are considered. Medium-deep voids should, nonetheless, be useful for survey self-calibration. Combined constraints from voids of different depth helps break degeneracies, such that background expansion, growth rate of structure, and power spectrum can be estimated fairly independently of each other. 

Combining parameter constraints from cluster and void abundances in future surveys could ideally constrain deviations from GR+$\Lambda$CDM on cosmological scales to percent level. The combination can improve the dark-energy/modified-gravity Figures of Merit a factor of 20 or more relative to individual abundances, and ideally a factor $600+$ relative to current cosmological data. This is due to clusters and voids having complementary redshift sensitivity to growth of structure vs. volume expansion, and voids probing the matter power spectrum more directly and across a wider range of scales than clusters do. Void surveys are sensitive to linear comoving scales up to a factor $2-4$ smaller than their limiting radius, and can cover the full range in scale from matter--radiation equality turnover $k_{\rm eq} \sim 0.01 \,h$Mpc$^{-1}$, through baryon acoustic peaks $k_{\rm BAO} \sim 0.06 \,h$Mpc$^{-1}$, to the cluster quasilinear regime $k_{\rm ql} \sim 0.1 \,h$Mpc$^{-1}$. The statistical power is independent of data from the cosmic microwave background, and hence can provide a precise and independent late-Universe probe of the power spectrum of density fluctuations (but cosmic microwave background data will improve constraints). 

Including additional statistics (e.g. correlation functions) and properties (e.g. measurements of cluster masses, void/cluster density profiles, gravitational lensing, ellipticities) of the void and cluster distributions should improve on this significantly. The ongoing development of void cosmology carries great potential to provide added value to current and future large-area surveys for constraining deviations from the cosmological concordance model, at low or no additional cost.

\onecolumngrid
\twocolumngrid

\begin{acknowledgments}
We thank Tom Kitching, Brice M\'enard and Johan Richard for helpful input, and the anonymous referees for good suggestions. MS thanks the Department of Physics \& Astronomy at the Johns Hopkins University for hospitality during the preparation of this work. M.S. was supported by the Olle Engkvist Foundation (Stiftelsen Olle Engkvist Byggm\"astare), the US-Sweden Fulbright Commission, the Helge Ax:son Johnson Foundation (Helge Ax:son Johnsons stiftelse), and the L\"angmanska Fund for Culture (L\"angmanska kulturfonden). J.S. was supported by European Research Council ``Ideas'' programme (Seventh Framework Programme) ERC Project No. 267117 Dark Matters (DARK) hosted by Universit\'e Pierre et Marie Curie (UPMC) - Paris 6.
\end{acknowledgments}


\begin{thebibliography}{28}
\expandafter\ifx\csname natexlab\endcsname\relax\def\natexlab#1{#1}\fi
\expandafter\ifx\csname bibnamefont\endcsname\relax
  \def\bibnamefont#1{#1}\fi
\expandafter\ifx\csname bibfnamefont\endcsname\relax
  \def\bibfnamefont#1{#1}\fi
\expandafter\ifx\csname citenamefont\endcsname\relax
  \def\citenamefont#1{#1}\fi
\expandafter\ifx\csname url\endcsname\relax
  \def\url#1{\texttt{#1}}\fi
\expandafter\ifx\csname urlprefix\endcsname\relax\def\urlprefix{URL }\fi
\providecommand{\bibinfo}[2]{#2}
\providecommand{\eprint}[2][]{\url{#2}}

\bibitem[{\citenamefont{{Rapetti} et~al.}(2013)\citenamefont{{Rapetti},
  {Blake}, {Allen}, {Mantz}, {Parkinson}, and {Beutler}}}]{2013MNRAS.432..973R}
\bibinfo{author}{\bibfnamefont{D.}~\bibnamefont{{Rapetti}}},
  \bibinfo{author}{\bibfnamefont{C.}~\bibnamefont{{Blake}}},
  \bibinfo{author}{\bibfnamefont{S.~W.} \bibnamefont{{Allen}}},
  \bibinfo{author}{\bibfnamefont{A.}~\bibnamefont{{Mantz}}},
  \bibinfo{author}{\bibfnamefont{D.}~\bibnamefont{{Parkinson}}},
  \bibnamefont{and}
  \bibinfo{author}{\bibfnamefont{F.}~\bibnamefont{{Beutler}}},
  \bibinfo{journal}{\mnras} \textbf{\bibinfo{volume}{432}},
  \bibinfo{pages}{973} (\bibinfo{year}{2013}), \eprint{1205.4679}.

\bibitem[{\citenamefont{{Xu}}(2013)}]{2013PhRvD..88h4032X}
\bibinfo{author}{\bibfnamefont{L.}~\bibnamefont{{Xu}}}, \bibinfo{journal}{\prd}
  \textbf{\bibinfo{volume}{88}}, \bibinfo{eid}{084032} (\bibinfo{year}{2013}),
  \eprint{1306.2683}.

\bibitem[{\citenamefont{{Pisani} et~al.}(2015)\citenamefont{{Pisani}, {Sutter},
  {Hamaus}, {Alizadeh}, {Biswas}, {Wandelt}, and
  {Hirata}}}]{2015arXiv150307690P}
\bibinfo{author}{\bibfnamefont{A.}~\bibnamefont{{Pisani}}},
  \bibinfo{author}{\bibfnamefont{P.~M.} \bibnamefont{{Sutter}}},
  \bibinfo{author}{\bibfnamefont{N.}~\bibnamefont{{Hamaus}}},
  \bibinfo{author}{\bibfnamefont{E.}~\bibnamefont{{Alizadeh}}},
  \bibinfo{author}{\bibfnamefont{R.}~\bibnamefont{{Biswas}}},
  \bibinfo{author}{\bibfnamefont{B.~D.} \bibnamefont{{Wandelt}}},
  \bibnamefont{and} \bibinfo{author}{\bibfnamefont{C.~M.}
  \bibnamefont{{Hirata}}}, \bibinfo{journal}{\prd}
  \textbf{\bibinfo{volume}{92}}, \bibinfo{eid}{083531} (\bibinfo{year}{2015}),
  \eprint{1503.07690}.

\bibitem[{\citenamefont{{Allen} et~al.}(2011)\citenamefont{{Allen}, {Evrard},
  and {Mantz}}}]{2011ARA&A..49..409A}
\bibinfo{author}{\bibfnamefont{S.~W.} \bibnamefont{{Allen}}},
  \bibinfo{author}{\bibfnamefont{A.~E.} \bibnamefont{{Evrard}}},
  \bibnamefont{and} \bibinfo{author}{\bibfnamefont{A.~B.}
  \bibnamefont{{Mantz}}}, \bibinfo{journal}{\araa}
  \textbf{\bibinfo{volume}{49}}, \bibinfo{pages}{409} (\bibinfo{year}{2011}),
  \eprint{1103.4829}.

\bibitem[{\citenamefont{{Blumenthal} et~al.}(1992)\citenamefont{{Blumenthal},
  {da Costa}, {Goldwirth}, {Lecar}, and {Piran}}}]{1992ApJ...388..234B}
\bibinfo{author}{\bibfnamefont{G.~R.} \bibnamefont{{Blumenthal}}},
  \bibinfo{author}{\bibfnamefont{L.~N.} \bibnamefont{{da Costa}}},
  \bibinfo{author}{\bibfnamefont{D.~S.} \bibnamefont{{Goldwirth}}},
  \bibinfo{author}{\bibfnamefont{M.}~\bibnamefont{{Lecar}}}, \bibnamefont{and}
  \bibinfo{author}{\bibfnamefont{T.}~\bibnamefont{{Piran}}},
  \bibinfo{journal}{\apj} \textbf{\bibinfo{volume}{388}}, \bibinfo{pages}{234}
  (\bibinfo{year}{1992}).

\bibitem[{\citenamefont{{Lam} et~al.}(2015)\citenamefont{{Lam}, {Clampitt},
  {Cai}, and {Li}}}]{2015MNRAS.450.3319L}
\bibinfo{author}{\bibfnamefont{T.~Y.} \bibnamefont{{Lam}}},
  \bibinfo{author}{\bibfnamefont{J.}~\bibnamefont{{Clampitt}}},
  \bibinfo{author}{\bibfnamefont{Y.-C.} \bibnamefont{{Cai}}}, \bibnamefont{and}
  \bibinfo{author}{\bibfnamefont{B.}~\bibnamefont{{Li}}},
  \bibinfo{journal}{\mnras} \textbf{\bibinfo{volume}{450}},
  \bibinfo{pages}{3319} (\bibinfo{year}{2015}), \eprint{1408.5338}.

\bibitem[{\citenamefont{{Brandbyge} et~al.}(2010)\citenamefont{{Brandbyge},
  {Hannestad}, {Haugb{\o}lle}, and {Wong}}}]{2010JCAP...09..014B}
\bibinfo{author}{\bibfnamefont{J.}~\bibnamefont{{Brandbyge}}},
  \bibinfo{author}{\bibfnamefont{S.}~\bibnamefont{{Hannestad}}},
  \bibinfo{author}{\bibfnamefont{T.}~\bibnamefont{{Haugb{\o}lle}}},
  \bibnamefont{and} \bibinfo{author}{\bibfnamefont{Y.~Y.~Y.}
  \bibnamefont{{Wong}}}, \bibinfo{journal}{\jcap} \textbf{\bibinfo{volume}{9}},
  \bibinfo{eid}{014} (\bibinfo{year}{2010}), \eprint{1004.4105}.

\bibitem[{\citenamefont{{Massara} et~al.}(2015)\citenamefont{{Massara},
  {Villaescusa-Navarro}, {Viel}, and {Sutter}}}]{2015arXiv150603088M}
\bibinfo{author}{\bibfnamefont{E.}~\bibnamefont{{Massara}}},
  \bibinfo{author}{\bibfnamefont{F.}~\bibnamefont{{Villaescusa-Navarro}}},
  \bibinfo{author}{\bibfnamefont{M.}~\bibnamefont{{Viel}}}, \bibnamefont{and}
  \bibinfo{author}{\bibfnamefont{P.~M.} \bibnamefont{{Sutter}}},
  \bibinfo{journal}{\jcap} \textbf{\bibinfo{volume}{11}}, \bibinfo{eid}{018}
  (\bibinfo{year}{2015}), \eprint{1506.03088}.

\bibitem[{\citenamefont{{Chongchitnan} and {Silk}}(2010)}]{2010ApJ...724..285C}
\bibinfo{author}{\bibfnamefont{S.}~\bibnamefont{{Chongchitnan}}}
  \bibnamefont{and} \bibinfo{author}{\bibfnamefont{J.}~\bibnamefont{{Silk}}},
  \bibinfo{journal}{\apj} \textbf{\bibinfo{volume}{724}}, \bibinfo{pages}{285}
  (\bibinfo{year}{2010}), \eprint{1007.1230}.

\bibitem[{\citenamefont{{Sahl{\'e}n} et~al.}(2016)\citenamefont{{Sahl{\'e}n},
  {Zubeld\'ia}, and {Silk}}}]{2016ApJ...820L...7S}
\bibinfo{author}{\bibfnamefont{M.}~\bibnamefont{{Sahl{\'e}n}}},
  \bibinfo{author}{\bibfnamefont{{\'I}.}~\bibnamefont{{Zubeld\'ia}}},
  \bibnamefont{and} \bibinfo{author}{\bibfnamefont{J.}~\bibnamefont{{Silk}}},
  \bibinfo{journal}{\apjl} \textbf{\bibinfo{volume}{820}}, \bibinfo{eid}{L7}
  (\bibinfo{year}{2016}), \eprint{1511.04075}.

\bibitem[{\citenamefont{{Laureijs} et~al.}(2011)\citenamefont{{Laureijs},
  {Amiaux}, {Arduini}, {Augu{\`e}res}, {Brinchmann}, {Cole}, {Cropper},
  {Dabin}, {Duvet}, {Ealet} et~al.}}]{2011arXiv1110.3193L}
\bibinfo{author}{\bibfnamefont{R.}~\bibnamefont{{Laureijs}}},
  \bibinfo{author}{\bibfnamefont{J.}~\bibnamefont{{Amiaux}}},
  \bibinfo{author}{\bibfnamefont{S.}~\bibnamefont{{Arduini}}},
  \bibinfo{author}{\bibfnamefont{J.~.} \bibnamefont{{Augu{\`e}res}}},
  \bibinfo{author}{\bibfnamefont{J.}~\bibnamefont{{Brinchmann}}},
  \bibinfo{author}{\bibfnamefont{R.}~\bibnamefont{{Cole}}},
  \bibinfo{author}{\bibfnamefont{M.}~\bibnamefont{{Cropper}}},
  \bibinfo{author}{\bibfnamefont{C.}~\bibnamefont{{Dabin}}},
  \bibinfo{author}{\bibfnamefont{L.}~\bibnamefont{{Duvet}}},
  \bibinfo{author}{\bibfnamefont{A.}~\bibnamefont{{Ealet}}},
  \bibnamefont{et~al.}, \bibinfo{journal}{ArXiv e-prints}
  (\bibinfo{year}{2011}), \eprint{1110.3193}.

\bibitem[{\citenamefont{de~Jong et~al.}(2014)\citenamefont{de~Jong, Barden,
  Bellido-Tirado, Brynnel, Chiappini, Depagne, Haynes, Johl, Phillips, Schnurr
  et~al.}}]{doi:10.1117/12.2055826}
\bibinfo{author}{\bibfnamefont{R.~S.} \bibnamefont{de~Jong}},
  \bibinfo{author}{\bibfnamefont{S.}~\bibnamefont{Barden}},
  \bibinfo{author}{\bibfnamefont{O.}~\bibnamefont{Bellido-Tirado}},
  \bibinfo{author}{\bibfnamefont{J.}~\bibnamefont{Brynnel}},
  \bibinfo{author}{\bibfnamefont{C.}~\bibnamefont{Chiappini}},
  \bibinfo{author}{\bibfnamefont{E.}~\bibnamefont{Depagne}},
  \bibinfo{author}{\bibfnamefont{R.}~\bibnamefont{Haynes}},
  \bibinfo{author}{\bibfnamefont{D.}~\bibnamefont{Johl}},
  \bibinfo{author}{\bibfnamefont{D.~P.} \bibnamefont{Phillips}},
  \bibinfo{author}{\bibfnamefont{O.}~\bibnamefont{Schnurr}},
  \bibnamefont{et~al.}, \emph{\bibinfo{title}{4most: 4-metre multi-object
  spectroscopic telescope}} (\bibinfo{year}{2014}),
  \urlprefix\url{http://dx.doi.org/10.1117/12.2055826}.

\bibitem[{\citenamefont{{Sartoris} et~al.}(2016)\citenamefont{{Sartoris},
  {Biviano}, {Fedeli}, {Bartlett}, {Borgani}, {Costanzi}, {Giocoli},
  {Moscardini}, {Weller}, {Ascaso} et~al.}}]{2015arXiv150502165S}
\bibinfo{author}{\bibfnamefont{B.}~\bibnamefont{{Sartoris}}},
  \bibinfo{author}{\bibfnamefont{A.}~\bibnamefont{{Biviano}}},
  \bibinfo{author}{\bibfnamefont{C.}~\bibnamefont{{Fedeli}}},
  \bibinfo{author}{\bibfnamefont{J.~G.} \bibnamefont{{Bartlett}}},
  \bibinfo{author}{\bibfnamefont{S.}~\bibnamefont{{Borgani}}},
  \bibinfo{author}{\bibfnamefont{M.}~\bibnamefont{{Costanzi}}},
  \bibinfo{author}{\bibfnamefont{C.}~\bibnamefont{{Giocoli}}},
  \bibinfo{author}{\bibfnamefont{L.}~\bibnamefont{{Moscardini}}},
  \bibinfo{author}{\bibfnamefont{J.}~\bibnamefont{{Weller}}},
  \bibinfo{author}{\bibfnamefont{B.}~\bibnamefont{{Ascaso}}},
  \bibnamefont{et~al.}, \bibinfo{journal}{\mnras}
  \textbf{\bibinfo{volume}{459}}, \bibinfo{pages}{1764} (\bibinfo{year}{2016}),
  \eprint{1505.02165}.

\bibitem[{\citenamefont{{Amendola} et~al.}(2016)\citenamefont{{Amendola},
  {Appleby}, {Avgoustidis}, {Bacon}, {Baker}, {Baldi}, {Bartolo}, {Blanchard},
  {Bonvin}, {Borgani} et~al.}}]{2016arXiv160600180A}
\bibinfo{author}{\bibfnamefont{L.}~\bibnamefont{{Amendola}}},
  \bibinfo{author}{\bibfnamefont{S.}~\bibnamefont{{Appleby}}},
  \bibinfo{author}{\bibfnamefont{A.}~\bibnamefont{{Avgoustidis}}},
  \bibinfo{author}{\bibfnamefont{D.}~\bibnamefont{{Bacon}}},
  \bibinfo{author}{\bibfnamefont{T.}~\bibnamefont{{Baker}}},
  \bibinfo{author}{\bibfnamefont{M.}~\bibnamefont{{Baldi}}},
  \bibinfo{author}{\bibfnamefont{N.}~\bibnamefont{{Bartolo}}},
  \bibinfo{author}{\bibfnamefont{A.}~\bibnamefont{{Blanchard}}},
  \bibinfo{author}{\bibfnamefont{C.}~\bibnamefont{{Bonvin}}},
  \bibinfo{author}{\bibfnamefont{S.}~\bibnamefont{{Borgani}}},
  \bibnamefont{et~al.}, \bibinfo{journal}{ArXiv e-prints}
  (\bibinfo{year}{2016}), \eprint{1606.00180}.

\bibitem[{\citenamefont{{4MOST Consortium}}()}]{4mostWP}
\bibinfo{author}{\bibnamefont{{4MOST Consortium}}}, \bibinfo{note}{in prep.}

\bibitem[{\citenamefont{{Sahl{\'e}n} et~al.}(2009)\citenamefont{{Sahl{\'e}n},
  {Viana}, {Liddle}, {Romer}, {Davidson}, {Hosmer}, {Lloyd-Davies}, {Sabirli},
  {Collins}, {Freeman} et~al.}}]{Sahlen2009}
\bibinfo{author}{\bibfnamefont{M.}~\bibnamefont{{Sahl{\'e}n}}},
  \bibinfo{author}{\bibfnamefont{P.~T.~P.} \bibnamefont{{Viana}}},
  \bibinfo{author}{\bibfnamefont{A.~R.} \bibnamefont{{Liddle}}},
  \bibinfo{author}{\bibfnamefont{A.~K.} \bibnamefont{{Romer}}},
  \bibinfo{author}{\bibfnamefont{M.}~\bibnamefont{{Davidson}}},
  \bibinfo{author}{\bibfnamefont{M.}~\bibnamefont{{Hosmer}}},
  \bibinfo{author}{\bibfnamefont{E.}~\bibnamefont{{Lloyd-Davies}}},
  \bibinfo{author}{\bibfnamefont{K.}~\bibnamefont{{Sabirli}}},
  \bibinfo{author}{\bibfnamefont{C.~A.} \bibnamefont{{Collins}}},
  \bibinfo{author}{\bibfnamefont{P.~E.} \bibnamefont{{Freeman}}},
  \bibnamefont{et~al.}, \bibinfo{journal}{\mnras}
  \textbf{\bibinfo{volume}{397}}, \bibinfo{pages}{577} (\bibinfo{year}{2009}),
  \eprint{0802.4462}.

\bibitem[{\citenamefont{{Linder}}(2005)}]{2005PhRvD..72d3529L}
\bibinfo{author}{\bibfnamefont{E.~V.} \bibnamefont{{Linder}}},
  \bibinfo{journal}{\prd} \textbf{\bibinfo{volume}{72}}, \bibinfo{eid}{043529}
  (\bibinfo{year}{2005}), \eprint{astro-ph/0507263}.

\bibitem[{\citenamefont{Di~Porto et~al.}(2012)\citenamefont{Di~Porto, Amendola,
  and Branchini}}]{DiPorto:2012ey}
\bibinfo{author}{\bibfnamefont{C.}~\bibnamefont{Di~Porto}},
  \bibinfo{author}{\bibfnamefont{L.}~\bibnamefont{Amendola}}, \bibnamefont{and}
  \bibinfo{author}{\bibfnamefont{E.}~\bibnamefont{Branchini}},
  \bibinfo{journal}{Mon. Not. Roy. Astron. Soc.}
  \textbf{\bibinfo{volume}{423}}, \bibinfo{pages}{L97} (\bibinfo{year}{2012}),
  \eprint{1201.2455}.

\bibitem[{\citenamefont{{Watson} et~al.}(2013)\citenamefont{{Watson}, {Iliev},
  {D'Aloisio}, {Knebe}, {Shapiro}, and {Yepes}}}]{Watson2013}
\bibinfo{author}{\bibfnamefont{W.~A.} \bibnamefont{{Watson}}},
  \bibinfo{author}{\bibfnamefont{I.~T.} \bibnamefont{{Iliev}}},
  \bibinfo{author}{\bibfnamefont{A.}~\bibnamefont{{D'Aloisio}}},
  \bibinfo{author}{\bibfnamefont{A.}~\bibnamefont{{Knebe}}},
  \bibinfo{author}{\bibfnamefont{P.~R.} \bibnamefont{{Shapiro}}},
  \bibnamefont{and} \bibinfo{author}{\bibfnamefont{G.}~\bibnamefont{{Yepes}}},
  \bibinfo{journal}{\mnras} \textbf{\bibinfo{volume}{433}},
  \bibinfo{pages}{1230} (\bibinfo{year}{2013}), \eprint{1212.0095}.

\bibitem[{\citenamefont{{Hu} and {Kravtsov}}(2003)}]{2003ApJ...584..702H}
\bibinfo{author}{\bibfnamefont{W.}~\bibnamefont{{Hu}}} \bibnamefont{and}
  \bibinfo{author}{\bibfnamefont{A.~V.} \bibnamefont{{Kravtsov}}},
  \bibinfo{journal}{\apj} \textbf{\bibinfo{volume}{584}}, \bibinfo{pages}{702}
  (\bibinfo{year}{2003}), \eprint{astro-ph/0203169}.

\bibitem[{\citenamefont{{Sheth} and {van de Weygaert}}(2004)}]{Sheth2004}
\bibinfo{author}{\bibfnamefont{R.~K.} \bibnamefont{{Sheth}}} \bibnamefont{and}
  \bibinfo{author}{\bibfnamefont{R.}~\bibnamefont{{van de Weygaert}}},
  \bibinfo{journal}{\mnras} \textbf{\bibinfo{volume}{350}},
  \bibinfo{pages}{517} (\bibinfo{year}{2004}), \eprint{astro-ph/0311260}.

\bibitem[{\citenamefont{{Jennings} et~al.}(2013)\citenamefont{{Jennings}, {Li},
  and {Hu}}}]{Jennings2013}
\bibinfo{author}{\bibfnamefont{E.}~\bibnamefont{{Jennings}}},
  \bibinfo{author}{\bibfnamefont{Y.}~\bibnamefont{{Li}}}, \bibnamefont{and}
  \bibinfo{author}{\bibfnamefont{W.}~\bibnamefont{{Hu}}},
  \bibinfo{journal}{\mnras} \textbf{\bibinfo{volume}{434}},
  \bibinfo{pages}{2167} (\bibinfo{year}{2013}), \eprint{1304.6087}.

\bibitem[{\citenamefont{{Leclercq} et~al.}(2015)\citenamefont{{Leclercq},
  {Jasche}, {Sutter}, {Hamaus}, and {Wandelt}}}]{2015JCAP...03..047L}
\bibinfo{author}{\bibfnamefont{F.}~\bibnamefont{{Leclercq}}},
  \bibinfo{author}{\bibfnamefont{J.}~\bibnamefont{{Jasche}}},
  \bibinfo{author}{\bibfnamefont{P.~M.} \bibnamefont{{Sutter}}},
  \bibinfo{author}{\bibfnamefont{N.}~\bibnamefont{{Hamaus}}}, \bibnamefont{and}
  \bibinfo{author}{\bibfnamefont{B.}~\bibnamefont{{Wandelt}}},
  \bibinfo{journal}{\jcap} \textbf{\bibinfo{volume}{3}}, \bibinfo{eid}{047}
  (\bibinfo{year}{2015}), \eprint{1410.0355}.

\bibitem[{\citenamefont{{Seljak} and
  {Zaldarriaga}}(1996)}]{1996ApJ...469..437S}
\bibinfo{author}{\bibfnamefont{U.}~\bibnamefont{{Seljak}}} \bibnamefont{and}
  \bibinfo{author}{\bibfnamefont{M.}~\bibnamefont{{Zaldarriaga}}},
  \bibinfo{journal}{\apj} \textbf{\bibinfo{volume}{469}}, \bibinfo{pages}{437}
  (\bibinfo{year}{1996}), \eprint{astro-ph/9603033}.

\bibitem[{\citenamefont{{Basilakos} and
  {Nesseris}}(2017)}]{2017PhRvD..96f3517B}
\bibinfo{author}{\bibfnamefont{S.}~\bibnamefont{{Basilakos}}} \bibnamefont{and}
  \bibinfo{author}{\bibfnamefont{S.}~\bibnamefont{{Nesseris}}},
  \bibinfo{journal}{\prd} \textbf{\bibinfo{volume}{96}}, \bibinfo{eid}{063517}
  (\bibinfo{year}{2017}), \eprint{1705.08797}.

\bibitem[{\citenamefont{{Levine} et~al.}(2002)\citenamefont{{Levine}, {Schulz},
  and {White}}}]{2002ApJ...577..569L}
\bibinfo{author}{\bibfnamefont{E.~S.} \bibnamefont{{Levine}}},
  \bibinfo{author}{\bibfnamefont{A.~E.} \bibnamefont{{Schulz}}},
  \bibnamefont{and} \bibinfo{author}{\bibfnamefont{M.}~\bibnamefont{{White}}},
  \bibinfo{journal}{\apj} \textbf{\bibinfo{volume}{577}}, \bibinfo{pages}{569}
  (\bibinfo{year}{2002}), \eprint{astro-ph/0204273}.

\bibitem[{\citenamefont{{Nadathur}}(2016)}]{2016MNRAS.461..358N}
\bibinfo{author}{\bibfnamefont{S.}~\bibnamefont{{Nadathur}}},
  \bibinfo{journal}{\mnras} \textbf{\bibinfo{volume}{461}},
  \bibinfo{pages}{358} (\bibinfo{year}{2016}), \eprint{1602.04752}.

\bibitem[{\citenamefont{{Sutter} et~al.}(2014)\citenamefont{{Sutter}, {Lavaux},
  {Wandelt}, {Weinberg}, {Warren}, and {Pisani}}}]{2014MNRAS.442.3127S}
\bibinfo{author}{\bibfnamefont{P.~M.} \bibnamefont{{Sutter}}},
  \bibinfo{author}{\bibfnamefont{G.}~\bibnamefont{{Lavaux}}},
  \bibinfo{author}{\bibfnamefont{B.~D.} \bibnamefont{{Wandelt}}},
  \bibinfo{author}{\bibfnamefont{D.~H.} \bibnamefont{{Weinberg}}},
  \bibinfo{author}{\bibfnamefont{M.~S.} \bibnamefont{{Warren}}},
  \bibnamefont{and} \bibinfo{author}{\bibfnamefont{A.}~\bibnamefont{{Pisani}}},
  \bibinfo{journal}{\mnras} \textbf{\bibinfo{volume}{442}},
  \bibinfo{pages}{3127} (\bibinfo{year}{2014}), \eprint{1310.7155}.

\end{thebibliography}
\end{document}